\documentclass[journal]{IEEEtran}
\usepackage{graphicx}
\usepackage{multirow,multicol}
\usepackage[linesnumbered,ruled,vlined]{algorithm2e}
\usepackage{amsmath}
\usepackage{setspace}
\usepackage{booktabs}
\usepackage{cellspace}%
\usepackage{makecell}
\usepackage{amsfonts}
\usepackage{url}
\usepackage{subfig}
\usepackage{cite}
\usepackage[font=small,labelfont=bf]{caption}
\usepackage{color, colortbl}
\definecolor{Gray}{gray}{0.9}
\ifCLASSINFOpdf
\else
\fi
\begin{document}
%
% paper title
% Titles are generally capitalized except for words such as a, an, and, as,
% at, but, by, for, in, nor, of, on, or, the, to and up, which are usually
% not capitalized unless they are the first or last word of the title.
% Linebreaks \\ can be used within to get better formatting as desired.
% Do not put math or special symbols in the title.
% \title{Bare Demo of IEEEtran.cls\\ for IEEE Journals}

% \title{NeXtformer: Re-designing Conformer and Multi-channel Speech Model for \\ Utterance and Continuous Speech Separation}

% \title{Dual-Path Multi-Channel Network for Speech \\ Separation, Dereverberation, and Enhancement}
\title{TF-CorrNet: Leveraging Spatial Correlation \\ for Continuous Speech Separation}
%
%
% author names and IEEE memberships
% note positions of commas and nonbreaking spaces ( ~ ) LaTeX will not break
% a structure at a ~ so this keeps an author's name from being broken across
% two lines.
% use \thanks{} to gain access to the first footnote area
% a separate \thanks must be used for each paragraph as LaTeX2e's \thanks
% was not built to handle multiple paragraphs
%
% Bon Hyeok Ku
\author{Ui-Hyeop Shin, Bon Hyeok Ku, and Hyung-Min Park, \IEEEmembership{Senior Member, IEEE}
\thanks{This work was supported in part by Institute of Information \& communications Technology Planning \& Evaluation (IITP) grant funded by the Korea government(MSIT)(RS-2022-II220989, Development of Artificial Intelligence Technology for Multi-speaker Dialog Modeling), and in part by Institute of Information \& communications Technology Planning \& Evaluation (IITP) grant funded by the Korea government(MSIT) (RS-2022-II220621, Development of artificial intelligence technology that provides dialog-based multi-modal explainability).}
% \thanks{This work was supported in part by Institute of Information \& communications Technology Planning \& Evaluation (IITP) grant funded by the Korea government (MSIT) (RS-2022-II220989, Development of Artificial Intelligence Technology for Multi-speaker Dialog Modeling), and in part by the National Research Foundation of Korea (NRF) and the Commercialization Promotion Agency for R\&D Outcomes (COMPA) grant funded by the Korea government (MSIT) (RS-2023-00237117).}
\thanks{The authors are with the Department of Electronic Engineering, Sogang University, Seoul 04107, Republic of Korea (e-mail: hpark@sogang.ac.kr).}}

\markboth{Journal of \LaTeX\ Class Files,~Vol.~14, No.~8, August~2015}%
{Shell \MakeLowercase{\textit{et al.}}: Bare Demo of IEEEtran.cls for IEEE Journals}
% The only time the second header will appear is for the odd numbered pages
% atfer the title page when using the twoside option.
% 
% *** Note that you probably will NOT want to include the author's ***
% *** name in the headers of peer review papers.                   ***
% You can use \ifCLASSOPTIONpeerreview for conditional compilation here if
% you desire.

% If you want to put a publisher's ID mark on the page you can do it like
% this:
%\IEEEpubid{0000--0000/00\$00.00~\copyright~2015 IEEE}
% Remember, if you use this you must call \IEEEpubidadjcol in the second
% column for its text to clear the IEEEpubid mark.

% use for special paper notices
%\IEEEspecialpapernotice{(Invited Paper)}

% make the title area
\maketitle

% As a general rule, do not put math, special symbols or citations
% in the abstract or keywords.
\begin{abstract}
% In this letter, we tackle continuous speech separation using correlation-based inputs.
% In this letter, we present TF-CorrNet for continuous speech separation using correlation-based inputs. 
In general, multi-channel source separation has utilized inter-microphone phase differences (IPDs) concatenated with magnitude information in time-frequency domain, or real and imaginary components stacked along the channel axis. However, the spatial information of a sound source is fundamentally contained in the ``differences" between microphones, specifically in the correlation between them, while the power of each microphone also provides valuable information about the source spectrum, which is why the magnitude is also included.
Therefore, we propose a network that directly leverages a correlation input with phase transform (PHAT)-$\beta$ to estimate the separation filter. 
In addition, the proposed TF-CorrNet processes the features alternately across time and frequency axes as a dual-path strategy in terms of spatial information. Furthermore, we add a spectral module to model source-related direct time-frequency patterns for improved speech separation. 
% In these processes, we introduce an efficient global and local Transformer as a processing unit to reduce computational complexity.
Experimental results demonstrate that the proposed TF-CorrNet effectively separates the speech sounds, showing high performance with a low computational cost in the LibriCSS dataset. 
% evaluation results of have released our on WHAMR! dataset code\footnote{\url{https://github.com/dmlguq456/PIT_CSS}} for future studies.
\end{abstract}

% Note that keywords are not normally used for peerreview papers.

% For peer review papers, you can put extra information on the cover
% page as needed:
% \ifCLASSOPTIONpeerreview
% \begin{center} \bfseries EDICS Category: 3-BBND \end{center}
% \fi
%
% For peerreview papers, this IEEEtran command inserts a page break and
% creates the second title. It will be ignored for other modes.
\IEEEpeerreviewmaketitle

\vspace{-.4mm}
\section{Introduction}

\IEEEPARstart{A}{utomatic} speech recognition (ASR) for continuous and overlapped speech is a challenging task for conversation transcription. To address this problem, continuous speech separation (CSS) is presented as a pre-processing step~\cite{Chen20_ICASSP, Chen_21_ICASSP} for overlapped speech recognition. For stable CSS, the use of multiple microphones significantly improves separation performance by leveraging spatial information compared to a single microphone. Conventionally, when training neural networks for multi-channel speech separation, inter-microphone phase differences (IPDs) are extracted and combined with magnitude spectra to provide spatial cues (e.g.~\cite{bahmaninezhad19_interspeech,Chen18_SLT,Chen_21_ICASSP,Yoshioka18_ICASSP,Wang19_TASLP,Gu20, Yoshioka22_ICASSP}). However, this approach may not yield the most optimal result, as it concatenates disparate spectral and spatial features, leaving room for further refinement. 
% The resulting large-dimensional input requires a significantly larger network to process effectively.
Meanwhile, neural beamforming methods have recently been proposed to replace conventional beamforming methods. They partially~\cite{Zhang21_ICASSP,Zhang22} or entirely~\cite{xu2021generalized, li21c_interspeech} replaced the beamforming filter estimation. Neural beamforming networks often directly process the target and noise instantaneous spatial covariance matrix estimates for each frequency bin independently.

More recently, it has often been preferred to use real and imaginary components of multi-channel STFT signals along the channel axis as an input to the network~\cite{Wang21_ICASSP, Tan22, TF_GridNet}. Although these methods offered significant performance improvements, they also required a complex model to learn the spatial features inherently included in inter-microphone information. In addition, it is known that these models were better suited for directly mapping output values than estimating filters because raw inputs are directly provided to the network~\cite{Wang21_ICASSP, TF_GridNet}. However, because the key spatial information is inherently captured in the inter-microphone correlations, leveraging these correlations can simplify the learning process. Moreover, filter estimation, rather than direct value mapping, can reliably preserve input signal components, as demonstrated by numerous array signal processing techniques~\cite{van1988beamforming, tashev2009sound}.

Therefore, we propose TF-CorrNet for continuous speech separation in multi-channel scenarios, including dereverberation and noise reduction. By directly utilizing inter-microphone correlations, TF-CorrNet simplifies the learning process by inherently capturing both spatial and spectral contexts without the need for explicitly concatenating disparate features or relying on raw signal components.
% TF-CorrNet uses inter-microphone correlation as the primary input, enabling the model to naturally learn and integrate both spatial and spectral information, simplifying the learning process. 
Also, to robustly capture spatial information, localization methods often apply a phase transform (PHAT) to the correlation value~\cite{Knapp76, Lee2020}. Furthermore, to control the balance of spectral and spatial information, we introduce PHAT-$\beta$~\cite{PHAT_beta} for correlations. We demonstrate that the filter estimation with correlation input is more effective in multi-channel separation compared to the other combinations, such as direct mapping with raw input.

Inspired by recent modeling in speech separation~\cite{TFPSNet, TF_GridNet} and enhancement~\cite{DPT_FSNet, CMGAN, MPSENet}, we employ dual-path modules to model the time and frequency sequence from the correlations. The temporal module learns spatial information that remains consistent along time axis for each frequency component that is analogous to neural beamforming~\cite{xu2021generalized, li21c_interspeech}, while the frequency module captures dependencies among frequency components. In addition, we incorporate a spectral module to directly model time-frequency patterns to improve speech separation as in the variance estimation of the conventional separation problem~\cite{Shin20, Kitamura16}.
Finally, to efficiently model these three modules with reduced computational complexity, we introduce an efficient local and global Transformer~\cite{shin24}. The proposed TF-CorrNet demonstrates improved performance on the LibriCSS dataset with significantly reduced computational cost and model size compared to previous approaches.

\begin{figure}
\footnotesize
\centering
\vspace{-1mm}
\includegraphics[width=0.91\columnwidth]{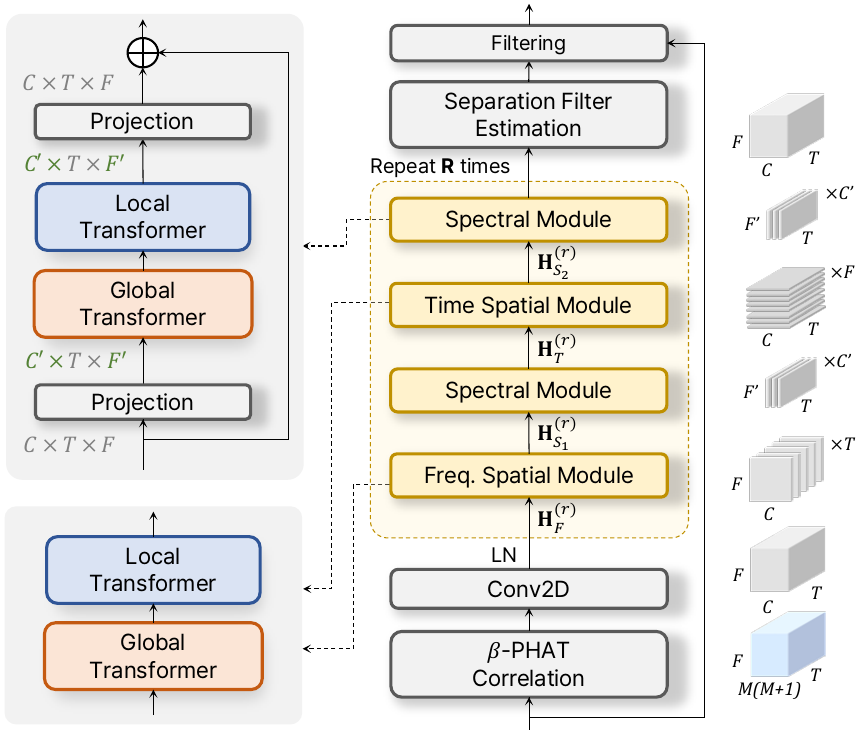}
\vspace{-1mm}
\caption{Overall structure of proposed TF-CorrNet.}
% \caption{{\bf Conventional multi-channel speech separation.} Concatenated input of reference magnitude and IPD is encoded to feature vector with dimensionality $G$ to be processed by separator blocks.}
\label{fig:BSS_update}
\vspace{-3mm}
\end{figure}

\section{TF-CorrNet for Speech Separation }

Let $M$-channel STFT observations be given as $\mathbf{x}_{tf} = [\hspace{-.4mm}{{X}_{\hspace{-.4mm}tf1\hspace{-.3mm}},...,\hspace{-.4mm}{X}_{tfM}}\hspace{-.4mm}]^T \hspace{-1.2mm}\in\hspace{-.8mm}\mathbb{C}^{M}\hspace{-.8mm},1\hspace{-1mm}\le\hspace{-1mm}t\hspace{-1mm}\le\hspace{-1mm}T, 1\hspace{-1mm}\le\hspace{-1mm} f\hspace{-1mm}\le\hspace{-1mm}F$, where $T$ and $F$ are the numbers of time frames and frequency bins, respectively. Then, we can formulate the input observation vector as $\mathbf{x}_{tf}\hspace{-1mm} = \hspace{-1mm}\sum\nolimits_{k=1}^K \hspace{-.5mm}\mathbf{h}_{k,f} S_{k,tf} \hspace{-.5mm}+\hspace{-.5mm} \mathbf{n}_{tf}$
% \vspace{-1mm}
% \begin{equation}
    % \mathbf{x}_{tf} = \sum\nolimits_{k=1}^K \mathbf{h}_{k,f} S_{k,tf} + \mathbf{n}_{tf},
% \end{equation}
where $S_{k,tf}$ and $\mathbf{h}_{k,tf}$ denote speech source and relative transfer function, respectively. $K$ is the number of speakers. $\mathbf{n}_{tf}$ is noise including late reverberations.
% In conventional methods, the input can be obtained by concatenating the reference magnitude $|{X}_{tf1}|$ and IPDs $\theta_{tf}^{(m,1)}=\hspace{-.5mm} \angle X_{tfm}\hspace{-.5mm} -\hspace{-.5mm} \angle X_{tf1}$. Or, the real and imaginary component is often stacked into the channel dimension.

\vspace{-1mm}

\subsection{Spatial correlations with PHAT-$\beta$}

Instead of concatenating the magnitude and IPD or stacking the real and imaginary components of the input $\mathbf{x}_{tf}$, the network input can be given by stacking the real and imaginary components of correlations as $\mathbf{z}_{tf}=[\text{Re}(\Phi_{tfmm'\hspace{-.5mm}}), \text{Im}({\Phi}_{tfmm'}\hspace{-.5mm})]_{1 \le m \le m'\le M} \hspace{-1mm}\in\hspace{-.5mm} \mathbb{R}^{M(M+1)}$, where ${\Phi}_{tfmm'}\hspace{-.5mm}=\hspace{-.5mm}{X}_{tfm}{X}^*_{tfm'}$ are spatial correlations.
Then, these input features include power information when $m=m'$, and otherwise, real and imaginary components retain IPDs as well as power information. However, power scales of correlations can lead to unstable training. Therefore, we utilized generalized PHAT-$\beta$~\cite{PHAT_beta} weighting as ${{\Phi}}_{tfmm'} \gets {{\Phi}_{tfmm'\hspace{-.5mm}} }/{|{\Phi}_{tfmm'\hspace{-.2mm}} |^\beta},0\hspace{-1.2mm}\le\hspace{-1mm} \beta\hspace{-1.2mm}\le\hspace{-1mm}1$.
% \vspace{-1mm}
% \begin{equation}
%     {{\Phi}}_{tfmm'} \gets {{\Phi}_{tfmm'} }/ {|{\Phi}_{tfmm'} |^\beta}, 0 \le \beta \le 1.
% \end{equation}
If $\beta\hspace{-.8mm}=\hspace{-.8mm}1$, it becomes the PHAT~\cite{Knapp76} and the network focuses on spatial clues without spectral patterns. 
% Conversely, when $\beta=0$, it could cause training to be unstable because the power values have much larger value range. 
% Furthermore, we set the value to be trained at each frequency bins.
Furthermore, we set the value to be trained at each frequency bin.
% In case of the IPD $\angle{{\Phi}}_{tfmm'}=\theta_{tf}^{(m,m')}$, we can also modulate as $\angle\tilde{{\Phi}}_{tfmm'} = \delta_f\angle\tilde{{\Phi}}_{tfmm'}$ where $\delta_f = \sigma(d_f)$ is trainable parameter. 
Then, the flattened PHAT-$\beta$ correlations $\mathbf{Z}\hspace{-1mm}\in\hspace{-.4mm}\mathbb{R}^{M(\hspace{-.2mm}M\hspace{-.2mm}+\hspace{-.2mm}1\hspace{-.2mm}) \times T \times F}$ are transformed into $\mathbf{H}_F\hspace{-1mm}\in \mathbb{R}^{C \times T \times F}$ by a 2d convolution layer followed by layer normalization (LN).

% \begin{figure*}
% \centering
% \subfloat[Global Transformer]{\includegraphics[width=0.28\textwidth]{Global_Transformer.pdf}}\hspace{7mm}
% \subfloat[Local Transformer]{\includegraphics[width=0.25\textwidth]{Local_Transformer.pdf}}\hspace{7mm}
% \subfloat[Efficient FFN]{\includegraphics[width=0.17\textwidth]{EFN.pdf}}
% \footnotesize
% \vspace{-1mm}
% \caption{{Block diagrams of (a) global Transformer and (b) local Transformer with their common (c) EFN module. In EGA and EFN, the downsampling and upsampling are denoted as $\downarrow$ and $\uparrow$.}}
% \vspace{-3mm}
% \label{fig:unit}
% \end{figure*}

\begin{figure}
\vspace{-4mm}
\hspace{-1.5mm}
\subfloat[Global Transformer]{\includegraphics[width=0.202\textwidth]{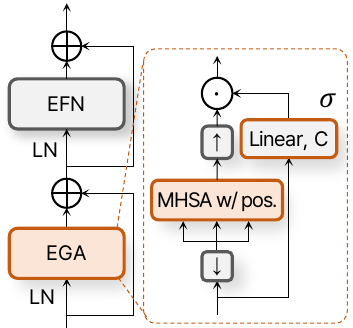}}
\subfloat[Local Transformer]{\includegraphics[width=0.173\textwidth]{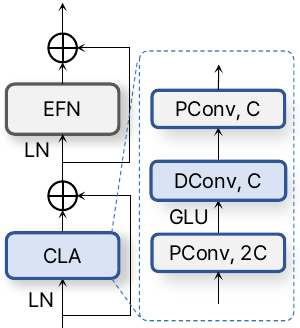}}
\subfloat[Efficient FFN]{\includegraphics[width=0.12\textwidth]{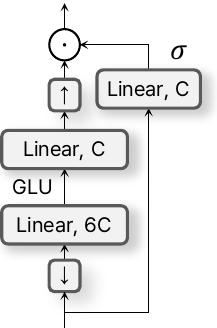}}
\footnotesize
\vspace{-1mm}
\caption{{Block diagrams of (a) global Transformer and (b) local Transformer with their common (c) EFN module. In EGA and EFN, the downsampling and upsampling are denoted as $\downarrow$ and $\uparrow$.}}
\vspace{-3mm}
\label{fig:unit}
\end{figure}

\vspace{-1mm}

\subsection{Time-Frequency Spatial Module}
A time-frequency feature with $C$ channels is processed through temporal and frequency modules, which are composed of global and local Transformers. In the frequency module of the $r$-th ($1\hspace{-.5mm}\le\hspace{-.5mm} r\hspace{-.5mm} \le\hspace{-.5mm} R$) stage, we consider the input features $\mathbf{H}_F^{(r)}\in \mathbb{R}^{T\times F \times C}$ as $T$ independent sequences with lengths of $F$. These sequences are processed by global-local Transformer blocks to strengthen the inter-frequency dependency inherently correlated by $\mathbf{h}_{k,f}$, whose phase values share a common factor of time differences, corresponding to the source direction. On the other hand, in the temporal module, we consider the input features $\mathbf{H}_T^{(r)}\hspace{-1mm}\in \mathbb{R}^{F\times T \times C}$ as $F$ independent sequences with lengths of $T$. From the temporal module, the network learns time-invariant spatial information by $\mathbf{h}_{k,f}$ across the frames at each frequency bin.

\vspace{-1mm}

\subsection{Spectral Module}
In addition to the spatial information, it is also important to properly model speech sources in conventional multi-channel speech processing such as beamforming~\cite{Cho19} and blind source separation~\cite{ono11, Kitamura16}. Therefore, we introduce spectral modules that learn the time-frequency patterns. In the spectral module, after projecting the input to the smaller dimension $C'\hspace{-1.2mm}<\hspace{-1mm}C$ by a linear layer, we consider the projected input features $\mathbf{H}_{S_1}^{(r)}\hspace{-.5mm},\mathbf{H}_{S_2}^{(r)}\hspace{-1mm}\in\hspace{-.5mm} \mathbb{R}^{C'\times T \times F}$ as $C'$ independent time-frequency representations. These spectral representations are again projected to the smaller dimension $F'\hspace{-1mm}<\hspace{-.8mm}F$ to model the latent spectral features. Then, the global-local Transformer learns the source-related spectral features on $C'$ independent sequences with lengths of $T$. Finally, the processed features are projected back to $C\hspace{-.5mm}\times\hspace{-.5mm}T\hspace{-.5mm}\times\hspace{-.5mm}F$ by two consecutive linear layers.

\vspace{-1mm}

\subsection{Filter Estimation}
The network outputs multi-input multi-output (MIMO) filters of $\mathbf{W}_{k,tf} \in \mathbb{C}^{M\times (2L+1)M},~1\le k \le K$ based on multi-tap filtering~\cite{DeepFiltering, Nakatani10} with the number of filter taps over frames $2L+1$ for improved separation in a reverberant mixture. The convolutional filters $\mathbf{W}_{k,tf}$ are applied to the input for separated output value $\mathbf{y}_{k,tf}=[Y_{k,tf1},...,Y_{k,tfM}]^T$  as
% \vspace{-0mm}
\begin{equation}
    \mathbf{y}_{k,tf} = \mathbf{W}_{k,tf}\tilde{\mathbf{x}}_{tf}\in \mathbb{C}^{M},
\end{equation}
where $\tilde{\mathbf{x}}_{tf}\hspace{-.8mm}=\hspace{-.8mm}[\mathbf{x}^T_{t-Lf},\hspace{-.5mm}...,\mathbf{x}^T_{tf},\hspace{-.5mm}...,\mathbf{x}^T_{t+Lf}]^T\hspace{-1.4mm}\in\hspace{-.8mm} \mathbb{C}^{(2L+1)M\hspace{-.8mm}}$. For filter estimation, the network splits the feature by a linear layer and reshape into $K\hspace{-.7mm}\times\hspace{-.7mm}C\hspace{-.7mm}\times\hspace{-.7mm}T\hspace{-.7mm}\times\hspace{-.7mm}F\hspace{-.5mm}$. Then, 2d convolution layer is applied to estimate $\mathbf{W}_{\hspace{-.5mm}k,tf}$.
For a multi-input single-output (MISO) case, the network can estimate $\mathbf{w}_{k,tf}\in\mathbb{C}^{1\times(2L+1)M}$.

\vspace{-1mm}
\section{Efficient Global and Local Transformer}

\subsection{Efficient Feed-forward Network }
For sequence modeling of each module, we introduce global and local Transformer~\cite{shin24} with an efficient feed-forward network (EFN) to reduce the computations as shown in Fig.~\ref{fig:unit}.
By performing the downsampling, the EFN can efficiently process input features. The downsampled feature is compensated by a gating mechanism. The vanilla FFN might be redundant because it performs equivalent operations in the time and frequency modules while processing $TF$-independent features. Therefore, the EFN enables the distinguished operations in the time and frequency modules. As an activation function for the hidden feature, we use a gated linear unit (GLU).

\vspace{-2mm}

\subsection{Global and Local Transformer}
\vspace{-.5mm}
While the global and local Transformer is analogous to the Conformer block~\cite{gulati20_interspeech}, we explicitly separate the role of capturing global and local contexts, respectively. In the global Transformer in Fig.~\ref{fig:unit}(a), the efficient global attention (EGA) module~\cite{shin24} is utilized to focus mainly on the global context to efficiently reduce computations. This EGA mechanism is similar to the EFN, in which the downsampled features are processed by multi-head self-attention (MHSA) and the output features are compensated for by a gating mechanism.
% \subsection{Efficient Local Transformer}
On the other hand, the local Transformer consists of convolutional local attention (CLA) and EFN. The CLA consists of 1d depth-wise convolution (DConv1D) and GLU to selectively capture local contexts between two point-wise convolution (PConv).

\begin{figure}
\footnotesize
\centering
\vspace{-1.5mm}
\includegraphics[width=0.95\columnwidth]{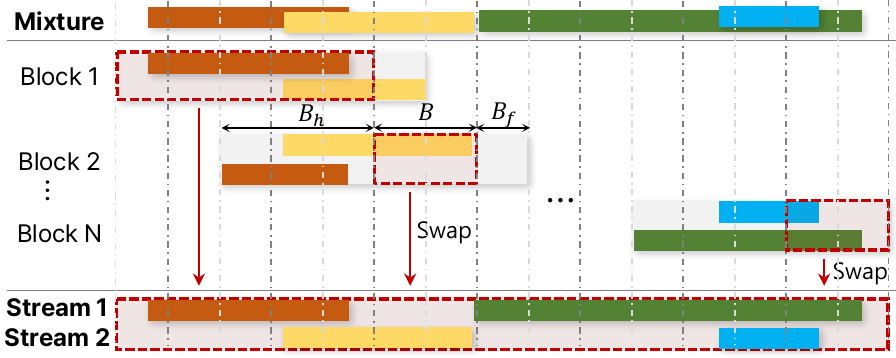}
% \vspace{-1mm}
\caption{{Illustration of CSS scheme. In our experiment, we set to chunk-size of $B_h = 1.2s, B = 0.8s$, and $B_f = 0.4s$, respectively.}}
% \caption{{\bf Conventional multi-channel speech separation.} Concatenated input of reference magnitude and IPD is encoded to feature vector with dimensionality $G$ to be processed by separator blocks.}
\label{fig:CSS}
\vspace{-2mm}
\end{figure}

\begin{figure}
\footnotesize
\centering
\includegraphics[width=0.7\columnwidth]{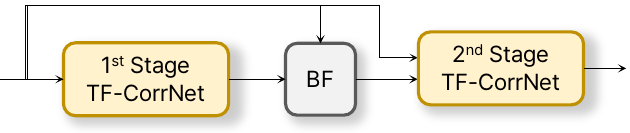}
\vspace{-1.5mm}
\caption{{Illustration of MIMO(MISO)-BF-MISO structure.}}
% \caption{{\bf Conventional multi-channel speech separation.} Concatenated input of reference magnitude and IPD is encoded to feature vector with dimensionality $G$ to be processed by separator blocks.}
\label{fig:2stage}
\vspace{-4mm}
\end{figure}

\section{Experiment}

\subsection{Dataset and Evaluation}
\vspace{-.5mm}
For evaluation, we used LibriCSS~\cite{Chen20_ICASSP}, a 10-hour 7-channel dataset recorded in a meeting-like scenario based on the Librispeech test set. The evaluation schemes are two ways: utterance-wise evaluation and continuous input evaluation. Because our focus is mainly on speech separation, we utilized an ASR model used in the original LibriCSS paper~\cite{Chen20_ICASSP} for evaluation. Following the previous studies~\cite{Chen20_ICASSP, Chen_21_ICASSP, Wang21_TASLP}, we used the same chunk-wise online processing and stitching with the same configurations as illustrated in Fig.~\ref{fig:CSS}. 

For improved separation results, we combined the separation model with a multi-channel Wiener filter and post filtering ~\cite{TF_GridNet, Taherian24_ICASSP} as shown in Figure~\ref{fig:2stage}. In each separation output, we concatenated the beamformed signal with the mixture as $M+1$ STFT information. This combined input was then fed into the enhancement model to output MISO filter $\tilde{\mathbf{w}}_{k,tf}\in\mathbb{C}^{1\times(2L+1)(M+1)}$. The enhancement model was trained by the same loss with the first stage network fixed. We denoted this model as MIMO(or MISO)-BF-MISO.
We trained the first stage network as MIMO when the localization-based stream merging was performed in CSS, otherwise, MISO.

\begin{table}
\footnotesize
\caption{Comparison of TF-CorrNet to TF-GridNet}
\vspace{-3mm}
\renewcommand{\tabcolsep}{3.5pt}
\def\arraystretch{1.0}
\begin{center}
\footnotesize
\begin{tabular}{lccccc}
% \Xhline{2.5\arrayrulewidth}
%\textbf{Method}
% \multirow{1}{*}{\hspace{-0.1mm}\textbf{Method}\hspace{-0.1mm}} & {\textbf{Param.}(M)$\downarrow$} & {\textbf{MACs}(G/s)$\downarrow$} & {\textbf{SDRi}(dB)$\uparrow$} & \multirow{1}{*}{\textbf{PESQ}$\uparrow$} & \textbf{STOI}$\uparrow$\\
\multirow{2}{*}{\hspace{-0.1mm}\textbf{Method}\hspace{-0.1mm}} & {\textbf{Param.}$\downarrow$} & {\textbf{MACs}$\downarrow$} & {\textbf{SDRi}$\uparrow$} & \multirow{2}{*}{\textbf{PESQ}$\uparrow$} & \multirow{2}{*}{\textbf{STOI}$\uparrow$}\\
&\textbf{(M)}&\textbf{(G/s)}&\textbf{(dB)}&\\
% &\textbf{(M)}&\textbf{(G/s)}&\textbf{(dB)}&\\
\Xhline{2.5\arrayrulewidth}
% t1 : 8.31 / 1.72
% t2 : 8.64 / 1.75

TF-GridNet & 5.6 & \hspace{-1.1mm}171.8 & 10.55 & 1.60 & 0.814\\
% TF-GridNet-\textit{small}& 1.5 & 43.9 &  & \\
% \hline
% \rowcolor{Gray}
TF-CorrNet (Conformer) & 4.2 & 85.5 & 11.75 & 1.81 & 0.866\\
\rowcolor{Gray}
TF-CorrNet & 5.1 & 44.5 & 11.38 & 1.75 & 0.857\\
% Proposed TF-CorrNet & 3.0 & 27.4 & 12.4 & 2.51\\
% \ fixed $\beta_f = 1$ & 3.0 & 27.4 & 11.6 & 2.36\\
% \ fixed $\beta_f = 0.5$ & 3.0 & 27.4 &12.3 & 2.48\\
% \ $|X_{tf1}|$, $\angle\Phi_{tfm1}$ instead of $\Phi_{tfmm'}\hspace{-5mm}$ & 3.0 & 27.2 & 9.6 & 2.21\\
% \ $X_{tfm}$ instead of $\Phi_{tfmm'}$ & 3.0 & 27.3 & 11.1 & 2.37\\
% \hline
% \ w/o spectral module ($R=8$)& 2.7 & 37.4 & 11.4 & 2.42\\
% \ Conformer as processing unit & 3.1 & 138.8 & 12.5 & 2.49\\

% \Xhline{2.5\arrayrulewidth}
\end{tabular}
\end{center}
\label{tab:sim}
\vspace{-4mm}
\end{table} 

% \begin{table}
% \footnotesize
% \caption{Comparison of SI-SNRi for WSJ0-2MIX and computations and parameter sizes.}
% \vspace{-3mm}
% \renewcommand{\tabcolsep}{6pt}
% \def\arraystretch{1.05}
% \begin{center}
% \footnotesize
% \begin{tabular}{lccc}
% % \Xhline{2.5\arrayrulewidth}
% %\textbf{Method}
% \multirow{1}{*}{\hspace{-0.1mm}\textbf{Network}\hspace{-0.1mm}} & \multirow{1}{*}{\hspace{-0.1mm}\textbf{Params.(M)}\hspace{-0.1mm}} & \multirow{1}{*}{\hspace{-0.1mm}\textbf{MACs(G/s)}\hspace{-0.1mm}} & {\textbf{SI-SNRi(dB)}}\\
% \Xhline{2.5\arrayrulewidth}
% % \hline
% TF-GridNet~\cite{TF_GridNet} & 8.4 & \ 72.4 & 22.2 \\
% TF-Locoformer(S)~\cite{Loco} & 5.0 & 150.2 & 22.0 \\
% \rowcolor{Gray}
% \textit{1ch} TF-CorrNet &  5.2 & \ 25.1 & 22.0\\
% % \ \ using diagonal and first row in ISCM & 5.4M & 9.7&\\
% % \  - $\tilde{\mathbf{C}}_{tfmm'}$ + ${X}_{tfm}$ \& $|X_{tfm}|$ & 5.4M & 9.7&\\
% % \  - trainable ${\beta}_f$ + fixed $\beta=1$ & 5.5M&10.8&\\
% % \  - trainable ${\beta}_f$ + fixed $\beta=0.5$ &5.5M&10.5&\\
% % \ \ w/o GRN layer & 5.4M & & \\
% % \  - trainable $\boldsymbol{\gamma}$  &5.5M&&\\
% % \ \ linear instead of attentive channel pool. &&&\\

% \end{tabular}
% \end{center}
% \label{tab:computation}
% \vspace{-4mm}
% \end{table} 

\begin{table*}
\footnotesize
\caption{WER(\%)$\downarrow$ results for utterance-wise and continuous evaluation on the LibriCSS}
\vspace{-3mm}
\renewcommand{\tabcolsep}{6.5pt}
\def\arraystretch{0.97}
\begin{center}
\footnotesize
\begin{tabular}{lccccccc|cccccc}
% \Xhline{2.5\arrayrulewidth}
%\textbf{Method} 
\multirow{2}{*}{\textbf{Method}}& \textbf{Stream} &\multicolumn{6}{c|}{\textbf{Utterance-wise}}&\multicolumn{6}{c}{\textbf{Continuous}} \\
& \textbf{Merging} &{\hspace{-0.1mm}0S\hspace{-0.7mm}} & {0L} & {10} & {20} & {30} & {40} &{\hspace{-0.1mm}0S\hspace{-0.7mm}} & {0L} & {10} & {20} & {30} & {40} \\
\Xhline{2.5\arrayrulewidth}
% \hline
\multicolumn{1}{l}{No Processing}&-&11.8&11.7&18.8&27.2&35.6&43.3&15.4&11.5&21.7&27.0&34.3&40.5\\
\multicolumn{1}{l}{Oracle Sound}&-&4.9&5.1&-&-&-&-&-&-&-&-&-&-\\
\hline
\multicolumn{1}{l}{BLSTM~\cite{Chen20_ICASSP}}&-&8.3&8.4&11.6&16.0&18.4&21.6&11.9&9.7&13.4&15.1&19.7&22.0\\
\multicolumn{1}{l}{Conformer~\cite{Chen_21_ICASSP}}&-&7.2&7.5&9.6&11.3&13.7&15.1&11.0&8.7&12.6&13.5&17.6&19.6\\
% \multicolumn{1}{l}{MISO(U-Net)~\cite{Wang21_TASLP}}&-&7.7&7.5&7.9&9.6&11.3&13.0&7.9&8.5&8.5&10.5&12.3&14.3\\
\multicolumn{1}{l}{MIMO (TF-GridNet)~\cite{Taherian24_ICASSP}}&-&7.5&7.4&7.3&8.3&9.6&10.3&9.2&12.2&9.9&10.1&11.9&12.2\\
% \hline
\multicolumn{1}{l}{MISO (U-Net)~\cite{Wang21_TASLP}}&SC&7.7&7.5&7.9&9.6&11.3&13.0&7.9&8.5&8.5&10.5&12.3&14.3\\
\multicolumn{1}{l}{MIMO (TF-GridNet)~\cite{Taherian24_ICASSP}}&LOC&5.8&6.4&6.7&7.9&9.5&10.3&8.4&9.1&9.4&10.0&11.3&12.0\\
% \rowcolor{Gray}
% \multicolumn{1}{l}{MISO (TF-CorrNet-\textit{s})}&-&5.9&6.1&6.7&7.8&9.6&10.1&8.2&8.9&9.2&9.5&10.5&11.6\\
\rowcolor{Gray}
\multicolumn{1}{l}{MISO (TF-CorrNet)}&-&5.8&{\bf 5.7}&6.4&{\bf 7.3}&8.6&{\bf 9.5}&7.6&8.1&8.3&9.1&10.1&10.9\\
\rowcolor{Gray}
\multicolumn{1}{l}{MIMO (TF-CorrNet)}&LOC&{\bf 5.6}&{5.8}&{\bf 6.3}&{7.4}&{\bf 8.6}&{9.7}  & {\bf 6.9}&{\bf 6.4}&{\bf 7.4}&{\bf 7.8}&{\bf 9.9}&{\bf 10.8}\\
% \rowcolor{Gray}
% \multicolumn{1}{l}{MIMO-BF (TF-CorrNet)}&LOC&{5.9}&{\bf 5.7}&{\bf 6.2}&{\bf 6.6}&{\bf 8.2}&{\bf 8.9}  & {\bf -}&{\bf -}&{\bf -}&{\bf -}&{\bf -}&{\bf -}\\
\hline
% \Xhline{2.5\arrayrulewidth}
\multicolumn{1}{l}{MISO-BF-MISO (TF-GridNet)~\cite{Taherian24_ICASSP}}&-&6.1&6.3&{5.9}&{6.1}&{6.7}&7.8&8.0&8.4&7.4&7.1&9.0&9.3\\
\multicolumn{1}{l}{MISO-BF-MISO (U-Net)~\cite{Wang21_TASLP}}&SC&5.8&5.8&{5.9}&{6.5}&7.7&8.3&7.7&7.6&7.4&8.4&9.7&11.3\\
\multicolumn{1}{l}{MIMO-BF-MISO (TF-GridNet)~\cite{Taherian24_ICASSP}}&LOC&{5.3}&5.7&{5.5}&{5.8}&{6.8}&7.1&6.8&6.8&{6.7}&6.9&8.4&9.0\\
\rowcolor{Gray}
\multicolumn{1}{l}{{MISO-BF-MISO (TF-CorrNet)}}&-&{5.8}&{5.7}&{5.7}&{5.9}&{6.6}&{7.3}&7.3&7.4&7.7&7.1&9.1&9.0\\
\rowcolor{Gray}
\multicolumn{1}{l}{{MIMO-BF-MISO (TF-CorrNet)}}&LOC&{\bf 5.3}&{\bf 5.5}&{\bf 5.5}&{\bf 5.7}&{\bf 6.4}&{\bf 6.7}&  {\bf 6.4}&{\bf 6.1}&  {\bf 6.2}&{\bf 6.2}&{\bf 7.4}&{\bf 7.7}\\

% \Xhline{2.5\arrayrulewidth}
\end{tabular}
\end{center}
\label{tab:LibriCSS}
\vspace{-2mm}
\end{table*}

\vspace{-2mm}
\subsection{Model Configuration and Training}
In the TF-CorrNet, the number of channels $C$ was set to $96$. For the spectral module, the projected channel dimension $C'$ was set to 16, and the spectral dimension $F'$ was set to $96$. In the EGA and the EFN, the downsampling factor was 4. The number of heads in the EGA was set to 4, and the kernel size of the DConv1D in the CLA to 65. $R$ is set to 4 considering the trade-off between the computational cost and performance.
% On similar model size and performance, computations are as shown in Table~\ref{tab:computation} which is much smaller compared to conventional TF-GridNet.
To train the network, we simulated reverberated 7-channel mixtures using the gpuRIR~\cite{gpurir_2021} with $T_{60} \in [0.2$s$,~0.6$s$]$ based on the Librispeech dataset~\cite{librispeech}. The sampling rate was 16kHz. In this case, we considered various mixing configurations, as in~\cite{Yoshioka18_ICASSP} with an average overlap ratio of about 50\%. We added spatially diffuse and random colored white noise with a signal-to-noise ratio (SNR) ranging from 0 to 20dB. 
The networks were trained with AdamW optimizer. The learning rate was set to 1e-4. If the validation loss did not decrease for two consecutive times, the learning rate was reduced by a factor of 0.8, and maximum 100 epochs were trained. In each epoch, 40,000 utterances, each of 2.4 seconds as shown in Fig.~\ref{fig:CSS}, are used with a batch size of 2.

When $\mathbf{Y}_k, \mathbf{S}_k \in \mathbb{C}^{M \times F\times T} $ were STFT values of estimated output $y_k$ and target source $s_k$, the proposed network was trained with a loss given as
% The proposed network was trained with the combination of magnitude loss and complex loss. When $\mathbf{Y}_k, \mathbf{S}_k \in \mathbb{C}^{M \times F\times T} $ were STFT values of estimated output $y_k$ and target source $s_k$, the loss was given as
\vspace{-1mm}
\begin{equation}
% \small
% \mathcal{L}_{Mag.} = \sum\nolimits_{k=1}^K \left\| \mathbf{Y}^{(a)}_{k}-\mathbf{S}^{(a)}_{k} \right\|_1, \hspace{28mm} \\
\mathcal{L}_{TF} \hspace{-.5mm}= \hspace{-.5mm}\sum_{k=1}^K\hspace{-1mm}\left(\left\| \mathbf{Y}^{(a)}_{k}\hspace{-1.5mm}-\mathbf{S}^{(a)}_{k}\hspace{-.3mm} \right\|_1 \hspace{-1.5mm}+\left\| \mathbf{Y}^{(r)}_{k}\hspace{-1.5mm}-\mathbf{S}^{(r)}_{k}\hspace{-.3mm}\right\|_1 \hspace{-1.5mm}+ \left\| \mathbf{Y}^{(i)}_{k}\hspace{-1.5mm}-\mathbf{S}^{(i)}_{k}\hspace{-.3mm}\right\|_1\right),
% \mathcal{L}_{RI} = \sum\nolimits_{k=1}^K\left(\left\| \mathbf{Y}^{(r)}_{k}-\mathbf{S}^{(r)}_{k}\right\|_1 + \left\| \mathbf{Y}^{(i)}_{k}-\mathbf{S}^{(i)}_{k}\right\|_1\right),
\end{equation}
% \vspace{-1mm}
where $\mathbf{Y}_k^{(a)}$, $\mathbf{Y}_k^{(r)}$, $\mathbf{Y}_k^{(i)}$ denote the magnitude, real, and imaginary components of $\mathbf{Y}_k$. $\|\cdot \|$ denotes L1-norm.
Moreover, a time-domain loss and a mixture-constraint loss~\cite{TF_GridNet} were additionally used, which was given as  $\mathcal{L}_{wav} = \sum\nolimits_{k=1}^K\left\| y_{k}-s_{k} \right\|_1$ and $\mathcal{L}_{MC} = \| \sum\nolimits_{k=1}^Ky_{k}-\sum\nolimits_{k=1}^Ks_{k} \|_1$, respectively.
% \begin{equation}
% \mathcal{L}_{wav} = \sum\nolimits_{k=1}^K\left\| y_{k}-s_{k} \right\|_1,
% \end{equation}
% \begin{equation}
% \mathcal{L}_{MC} = \left\| \sum\nolimits_{k=1}^Ky_{k}-\sum\nolimits_{k=1}^Ks_{k} \right\|_1.
% \end{equation}
Therefore, the final loss was given as $\mathcal{L} = \mathcal{L}_{TF} + \mathcal{L}_{wav} + \mathcal{L}_{MC}$ with permutation invariant training (PIT)~\cite{uPIT} to solve permutation ambiguity. % We did not considered weighting terms in loss because all the loss terms have similar scales.
In addition, we used the direct signal as the target, which means that the networks were trained to perform separation, dereverberation, and denoising simultaneously.

\subsection{Comparison on the Simulation Data}
\label{sec:comp_sim}
In Table~\ref{tab:sim}, we compared our proposed TF-CorrNet to the TF-GridNet \cite{TF_GridNet} using signal-to-distortion ratio improvement (SDRi)~\cite{1643671}, perceptual evaluation of speech quality (PESQ), short-time objective intelligibility (STOI) on simulated test mixtures, which were similarly generated to the simulated training set using the Librispeech test set. TF-GridNet follows the same configuration as in~\cite{Taherian24_ICASSP}. Compared to TF-GridNet, TF-CorrNet with Conformer achieved better performance while reducing computational costs. Furthermore, replacing Conformer with global-local Transformer as the unit block nearly halved the computational requirements, which showed the efficiency of the proposed model with the global-local Transformer.

\vspace{-1mm}
\subsection{Results on LibriCSS}

In Table~\ref{tab:LibriCSS}, our proposed method was compared with the recently introduced TF-GridNet-based approach~\cite{TF_GridNet, Taherian24_ICASSP} which has garnered attention for its good performance. The evaluation was further divided into two main categories: the MISO (or MIMO) models and the MISO (or MIMO)-BF-MISO models.
In the first group, our proposed TF-CorrNet demonstrated better performance than to the real-valued mask estimation models and the TF-GridNet-based approach, thereby validating the robustness of our method in terms of separation stability.
In the second group, compared to existing methods that integrate speaker counting (SC)~\cite{Wang21_ICASSP} or localization (LOC)~\cite{Taherian24_ICASSP} for stream merging, our approach already showed comparable performance without stream merging by effectively separating sources. Furthermore, when the proposed model was extended to MIMO with the integration of localization-based stream merging, we observed slight performance improvements. This suggests that our model is capable of reliably estimating the output phase for localization compared to conventional mapping methods.

Moreover, when the model was extended to a 2-stage system, it demonstrated significantly improved performance, benefiting from the stable results of beamforming and the additional post-enhancement network. Also, the proposed MIMO-BF-MISO method successfully separated sources without relying on an independent stream merging technique, achieving state-of-the-art performance on the LibriCSS dataset through localization-based stream merging. Notably, the proposed MISO and MISO-BF-MISO without stream merging still showed impressive separation performance, which suggests that the models perform stable separations with input correlations and filter estimation for outputs.

\subsection{Ablation Study}
For ablation study, we evaluated on the same configurations as Subsection~\ref{sec:comp_sim}.
In Table~\ref{tab:abl}, we experimented the case using the correlation value with or without PHAT weighting, which corresponded to the beta value of 1 or 0, respectively. Setting beta to 1 resulted in a more noticeable decline in performance as the network relied solely on spatial information, losing spectral details.
Using the magnitude of reference mic. $|X_{tf1}|$ and IPD as in conventional methods significantly reduced performance. Similarly, replacing the correlation with direct input signals also led to decreased performance.
The removal of the spectral module resulted in performance drop despite using larger $R$ with increased computational loads, highlighting the necessity of our proposed spectral module. 
% The last row showed that while using a Conformer block improved the results, it doubled the computational cost.

In Table~\ref{tab:filter_vs_mapping}, we evaluated four combinations of input types (correlation or raw input) and output types (filtering or mapping). Using raw input instead of correlation was less effective in the filtering approach. Furthermore, employing mapping instead of filtering with correlation-based input was found to be entirely unfeasible as the input phases to be enhanced directly are not available. On the other hand, using raw input with mapping did not result in a noticeable performance difference compared to its combination with filtering. Among these combinations, the best performance was achieved when correlation-based input was paired with filtering.

\begin{table}
\footnotesize
\caption{Ablation study for the proposed TF-CorrNet}
\vspace{-3mm}
\renewcommand{\tabcolsep}{2.7pt}
\def\arraystretch{1.0}
\begin{center}
\footnotesize
\begin{tabular}{lcccc}
% \Xhline{2.5\arrayrulewidth}
%\textbf{Method}
\multirow{2}{*}{\hspace{-0.1mm}\textbf{Method}\hspace{-0.1mm}} & {\textbf{Param.}$\downarrow$} & {\textbf{MACs}$\downarrow$} & {\textbf{SDRi}$\uparrow$} & \multirow{2}{*}{\textbf{PESQ}$\uparrow$}\\
&\textbf{(M)}&\textbf{(G/s)}&\textbf{(dB)}&\\
\Xhline{2.5\arrayrulewidth}
% t1 : 8.31 / 1.72
% t2 : 8.64 / 1.75

% TF-GridNet & 5.6 & \hspace{-1.1mm}171.8 & \hspace{-1.1mm}10.90 & 1.99\\
% \hline
\rowcolor{Gray}
Proposed TF-CorrNet & 5.1 & 44.5 & 11.38 & 1.75\\
% \hline
\ fixed $\beta_f = 0$ (no PHAT weighting) & 5.1 & 44.5 & 10.97 & 1.68\\
\ fixed $\beta_f = 1$ (PHAT weighting) & 5.1 & 44.5 & \hspace{1.2mm}9.77 & 1.55\\
% \ fixed $\beta_f = 0.5$ (PHAT-$\beta$ weighting) & 5.09 & 44.49 & 8.49 & 1.70\\
\hline
\ $|X_{tf1}|$, $\angle\Phi_{tfm1}$ instead of $\Phi_{tfmm'}\hspace{-2mm}$ & 5.1 & 43.7 & \hspace{1.2mm}8.52 & 1.41\\
\ $X_{tfm}$ instead of $\Phi_{tfmm'}$ & 5.1 & 43.9 & \hspace{1.2mm}9.05 & 1.46\\
\hline
\ w/o spectral module ($R=5$)& 3.0& 47.2 & 10.44 & 1.59\\
\ w/o spectral module ($R=6$)& 3.5& 54.7 & 11.08 & 1.65\\
% \hline
% \ Conformer as processing unit & 4.2 & 85.5 & 8.91 & 1.81\\

% Proposed TF-CorrNet & 3.0 & 27.4 & 12.4 & 2.51\\
% \ fixed $\beta_f = 1$ & 3.0 & 27.4 & 11.6 & 2.36\\
% \ fixed $\beta_f = 0.5$ & 3.0 & 27.4 &12.3 & 2.48\\
% \ $|X_{tf1}|$, $\angle\Phi_{tfm1}$ instead of $\Phi_{tfmm'}\hspace{-5mm}$ & 3.0 & 27.2 & 9.6 & 2.21\\
% \ $X_{tfm}$ instead of $\Phi_{tfmm'}$ & 3.0 & 27.3 & 11.1 & 2.37\\
% \hline
% \ w/o spectral module ($R=8$)& 2.7 & 37.4 & 11.4 & 2.42\\
% \ Conformer as processing unit & 3.1 & 138.8 & 12.5 & 2.49\\

% \Xhline{2.5\arrayrulewidth}
\end{tabular}
\end{center}
\label{tab:abl}
\vspace{-1mm}
\end{table}

\begin{table}
\footnotesize
\caption{Evaluation on various input-output configuration}
\vspace{-3mm}
\renewcommand{\tabcolsep}{7pt}
\def\arraystretch{1.0}
\begin{center}
\footnotesize
\begin{tabular}{cccc}
\multirow{1}{*}{\hspace{-0.1mm}\textbf{Input}\hspace{-0.1mm}} & \multirow{1}{*}{\hspace{-0.1mm}\textbf{Output}\hspace{-0.1mm}} & {\textbf{SDRi(dB)}$\uparrow$} & \multirow{1}{*}{\textbf{PESQ}$\uparrow$}\\
\Xhline{2.5\arrayrulewidth}
\rowcolor{Gray}
correlation $\Phi_{tfmm'}$ & Filtering &11.38&1.75\\ 
raw input $X_{tfm}$ & Filtering &\hspace{1.2mm}9.05&1.46\\ 
correlation $\Phi_{tfmm'}$ & Mapping & \hspace{0.5mm}-7.64 & 1.14\\
raw input $X_{tfm}$ & Mapping & \hspace{1.2mm}9.16 & 1.43  \\
% correlation $\Phi_{tfmm'}$ & Filtering &12.4&2.51\\ 
% raw input $X_{tfm}$ & Filtering &11.1&2.37\\ 
% correlation $\Phi_{tfmm'}$ & Mapping &6.8&2.12\\
% raw input $X_{tfm}$ & Mapping &11.8&2.33\\

\end{tabular}
\end{center}
\label{tab:filter_vs_mapping}
\vspace{-4mm}
\end{table}

\section{Conclusion}
We presented a TF-CorrNet for multi-channel speech separation. With correlation input and filter estimation, the network effectively performs separation. In addition, the proposed time-frequency spatial and spectral modules effectively process the feature for the multi-channel signals. Experimental results showed the effectiveness of TF-CorrNet in CSS, achieving state-of-the-art performance on the LibriCSS dataset.

% if have a single appendix:
%\appendix[Proof of the Zonklar Equations]
% or
%\appendix  % for no appendix heading
% do not use \section anymore atfer \appendix, only \section*
% is possibly needed

% use appendices with more than one appendix
% then use \section to start each appendix
% you must declare a \section before using any
% \subsection or using \label (\appendices by itself
% starts a section numbered zero.)
%

% \appendices
% \section{Proof of the First Zonklar Equation}
% Appendix one text goes here.

% you can choose not to have a title for an appendix
% if you want by leaving the argument blank

% use section* for acknowledgment
\pagebreak

% \section*{Acknowledgment}

% The authors would like to thank...

% Can use something like this to put references on a page
% by themselves when using endfloat and the captionsoff option.
\ifCLASSOPTIONcaptionsoff
  \newpage
\fi

% trigger a \newpage just before the given reference
% number - used to balance the columns on the last page
% adjust value as needed - may need to be readjusted if
% the document is modified later
%\IEEEtriggeratref{8}
% The "triggered" command can be changed if desired:
%\IEEEtriggercmd{\enlargethispage{-5in}}

% references section

% can use a bibliography generated by BibTeX as a .bbl file
% BibTeX documentation can be easily obtained at:
% http://mirror.ctan.org/biblio/bibtex/contrib/doc/
% The IEEEtran BibTeX style support page is at:
% http://www.michaelshell.org/tex/ieeetran/bibtex/
\bibliographystyle{IEEEtran}
% argument is your BibTeX string definitions and bibliography database(s)
%\bibliography{IEEEabrv,../bib/paper}
\setstretch{1.03}
\bibliography{bibtex/bib/IEEEabrv, reference}
\end{document}